\documentstyle{article}
\begin{document}

{\large
{

\centerline{\bf {THE ANOMALY-FREE QUANTIZATION }}
\centerline{\bf {OF TWO-DIMENSIONAL RELATIVISTIC STRING. I }}
\centerline{ }

\centerline{ \it S.N.Vergeles.\footnote{e-mail:Vergeles@itp.ac.ru}}
\vspace{3mm}
\centerline { \it {L. D. Landau Institute of Theoretical Physics,
Russian Academy of Sciences.}}
\centerline { \it {142432 Chernogolovka, Russia.}}
\centerline{ }
\centerline{Submitted December 11, 1997
 Zh. \'{E}ksp. Teor. Fiz. {\bf 113}, (1998) 1566--1578 }

\centerline { }
\centerline { }

\centerline{\bf {Abstract.}}
\centerline{ }

\parbox[b]{135mm}{
\hspace{5mm}
{An anomaly-free quantum theory of a relativistic string is constructed in
two-dimensional space-time. The states of the string are found to be similar
to the states of a massless chiral quantum particle. This result is obtained
by generalizing the concept of an ``operator'' in quantum field theory.
}}

\centerline{ }
\centerline {\bf {1. Introduction}}
\centerline { }

It has recently been asserted in a number of works (see, for example, Refs.
[1] and [2]) that anomaly-free quantization
of some models of two-dimensional gravity
is possible. Specifically, Ref.
[1]
examined a model of two-dimensional gravity Ref. [3] which in certain
variables was described by the same constraints as those
describing a relativistic bosonic string in two-dimensional space-time:
$$
{\cal E}=-{\cal E}_0+{\cal E}_1\approx 0\,,
$$
$$
{\cal E}_0=\frac{1}{2}\,\left(\,(\pi_0)^2+({r^0}^{\prime})^2\right)\,,
 \qquad {\cal E}_1=\frac{1}{2}\,\left(\,(\pi_1)^2+({r^1}^{\prime})^2\right)\,,
   \eqno(1.1a)
$$
$$
{\cal P}={r^{a}}^{\prime}\,\pi_a={r^0}^{\prime}\,\pi_0+
{r^1}^{\prime}\,\pi_1\approx 0  \eqno(1.1b)
$$
Dimensionless quantities are employed. Here $r^a(x)$ and $\pi _a(x),$
$a=0,\,1,$ are canonically conjugated fields on a one-dimensional manifold,
so that the nonzero commutation relations are
$$
[r^a(x), \, \pi _ b (y) \,] = i \,\delta ^ a _ b \,\delta (x-y) \eqno(1.2)
$$
A prime or overdot signifies a derivative $\partial /\partial x\,$ or
$\partial /\partial t$, respectively.

The ground state of the theory is determined at this stage of quantization.
This makes it possible to perform normal ordering of the operator in the
constraints (1.1). The determined normal ordering in the constraints can
lead in turn to anomalies in the commutators of the constraints. These
anomalies partially destroy the weak equalities (1.1). To determine the
ground state, the fields $r^a$ and $\pi _a$ are expanded in the modes that
arise when solving the Heisenberg equations
$$
i\, {\dot {r}}^a=[\, r^a,\,{\cal H}\,] \, , \qquad
i\, {\dot {\pi}}_a=[\, \pi_a, \,{\cal H}\,] \, ,
$$
$$
{\cal H} = \int \, dx \, {\cal E} \eqno (1.3)
$$
The solutions of Eqs. (1.2) and (1.3) can be written in the form
$$
r^a(t, \, x) = \int \,\frac {dk} {2 \,\pi} \, \frac {1} {\sqrt {2 | k |}} \,
\left \{c ^ a _ k \, e ^ {-i \, (| k | \, t-k \, x)} +
c^ {a +} _ k \, e ^ {i \, (| k | \, t-k \, x)} \, \right\} \,,
$$
$$
\pi^a (t, \, x) = -i \,\int \,\frac {dk} {2 \,\pi} \,
 \sqrt {\frac {| k |} {2}} \,
\left \{c^a_k \, e^{-i \, (| k | \, t-k \, x)} -
c^{a +}_k \, e ^ {i \, (| k | \, t-k \, x)} \, \right \} \,,
$$
$$
[\, c ^ a _ k, \, c ^ {b +} _ p \,] = 2 \,\pi \,\eta ^ {ab} \, \delta (k-p) \, ,
 \qquad [\, c ^ a _ k, \, c ^ b _ p \,] = 0 \eqno(1.4)
$$
Here \ $ \eta ^ {ab} $ \ (below - \ $ \eta ^ {\mu\nu}\,) = diag (-1,1) $.
We also  have the following commutational relations
$$
 [{\cal H},\, c^a_k\,] = - | k | \, c^a_k \,, \qquad
 [\,{\cal H}, \, c^{a+}_k \,] = |k| \, c^{a+}_k \eqno(1.5)
 $$

In conventional quantization the operators $c_k^a$ are considered as
annihilation
operators, while their hermitian conjugated operators $c_k^{a+}$
are considered as creation
operators. The ground state $\left| 0\right\rangle $ satisfies the
conditions
$$
c^a_k \,\vert \, 0 \,\rangle = 0 \eqno (1.6)
$$
Normal ordering of the operators $(c_k^{a+},c_k^a)$ in the quantities (1.1)
means that the creation operators stand to the left of all annihilation
operators.

Let us consider the state
$$
\vert \, k, \, a \,\rangle = c ^ {a +} _ k \, | \, 0 \,\rangle \eqno (1,7)
$$
It follows immediately from the commutation relations (1.5) that
$$
{\cal H} \, \vert \, k, \, a \,\rangle =
( | k | + E _ 0 \,) \,\vert \, k, \, a \,\rangle \,, \eqno (1.8)
$$
where $E_0$ is the value of the operator ${\cal H}$ for the ground state.
The equality (1.8) means that the operator ${\cal H}$ is
positive-definite.

In consequence of Eqs. (1.4) and (1.6) we have for the scalar product of the
vectors (1.7)
$$
\langle \, k, \, a \, | \, p, \, b \,\rangle = 2 \,\pi \,\eta ^ {ab} \, \delta (k-p) \eqno (1.9)
$$
Hence it is seen that the metric in the whole state space is indefinite.

Next, let us calculate the commutator $[{\cal E},\, {\cal P}]$. According to
Eq. (1.1) it can be represented as a sum of two terms
$$
[\,{\cal E}(x),\,{\cal P}(y)\,]=
-[\,{\cal E}_0(x),\,{r^0}^{\prime}\,\pi_0(y)\,]+
[\,{\cal E}_1(x),\,{r^1}^{\prime}\,\pi_1(x)\,]   \eqno(1.10)
$$
In consequence of Eq. (1.2), both commutators on the right-hand side of Eq.
(1.10) are identical up to a change of the index $a$. These commutators
are proportional (up to the ordering) to the quantities ${\cal E}_0$ and
${\cal E}_1$, respectively. As is well known, the normal ordering of the
operators in these commutators leads to anomalies.

Indeed, it follows from the commutation relations (1.4) that the
correspondences $c_k^0\leftrightarrow c_k^{1+}$ and $c_k^{0+}\leftrightarrow
c_k^1$ establish an isomorphism of the Heisenberg algebras $H_0$ and $H_1$,
whose generators are $(c_k^0,$ $c_k^{0+})$ and $(c_k^{1+},$ $c_k^1)$,
respectively. In this case the normal ordering of the operators in the
algebra $H_1$ is transferred by the indicated isomorphism into antinormal
ordering in the algebra $H_0$.

  As is known in such case the normal and antinormal
 orderings result in anomalies differing only in sign.
Therefore the contribution of the first commutator on the
right-hand side of the Eq. (1.10) to the anomaly will be $-A$ and that of
the second will be $A$. But, since a minus sign stands in front of the first
commutator in Eq. (1.10), the anomaly in Eq. (1.10) equals $-(-A)+A=2A$.

Let us now examine the problem from a different point of view.

In Ref. [1]
it is asserted that in the present theory the positive-definiteness
condition (1.8) for the operator ${\cal H}$ is not necessary. The initial
requirement of the theory is satisfaction of the weak equalities (1.1).
Therefore we have the right to reject the quantization conditions (1.6) and
replace them with the conditions
$$
c^{0 +} _ k \,\vert \, 0 \,\rangle = 0 \,, \qquad
 c^ 1 _ k \,\vert \, 0 \,\rangle = 0 \eqno (1.11)
$$

Under the quantization conditions (1.11) the basis of the whole Fock space of
the theory consists of vectors of the type
$$
c^ 0 _ {k _ 1} \, \ldots \, c ^ 0 _ {k _ m} \, c ^ {1 +} _ {p _ 1} \,
\ldots \, c ^ {1 +} _ {p _ n} \, \vert \, 0 \,\rangle  \eqno(1.12)
$$

It follows from the commutation relations (1.4) that the scalar product of
the states (1.12) is positive-definite. Moreover,
{\it {there is no anomaly in the
operator algebra (1.1).}}

Indeed, under the conditions (1.11) normal ordering consists of arranging
the operators $(c_k^0,\,c_k^{1+})$ to the left of all operators $(c_k^{0+},$
$c_k^1)$. This means that normal ordering occurs in both Heisenberg algebras
$H_0$ and $H_1$. Now, with normal ordering both commutators in (1.10) make
the same contribution, equal to $A,$ to the anomaly. Taking account of the
minus sign in front of the first commutator on the right-hand side of the
equality (1.10), the total anomaly is $-A+A=0$.

The absence of an anomaly in the operator algebra $({\cal E},$ ${\cal P})\,$
makes it possible to satisfy all weak equalities ${\cal E}\approx 0$ and
${\cal P}\approx 0$. Two physical states which the operators ${\cal E}$ and
${\cal P}$ annihilate are presented in Ref. [1]:
$$
\Psi _ {gravity} (r ^ a) = \exp\pm\frac {i} {2} \, \int \, dx \,\varepsilon _ {ab} \,
r^a \, {r ^ b} ^ {\prime}
$$

In the present paper we shall likewise reexamine the quantization conditions
for a relativistic string in two-dimensional space-time. In so doing, we
shall determine the space of physical states with a positive-definite scalar
product. The nonphysical states are not studied in the theory. The physical
states annihilate all  first class constraints, i.e. all Virasoro
operators. The physical states are characterized by a continuous parameter,
which has the meaning of momentum. However, in our theory not all dynamical
variables are linear operators in the space of physical states. In the
proposed theory the states of a relativistic string in two-dimensional
space-time are found to be identical to the states of a massless chiral
particle.

\centerline{ }
\centerline{\bf {2. Relativistic bosonic string}}
\centerline{\bf {in two-dimensional space - time}}
\centerline{ }

Let $X^\mu ,$ $\mu =0,\,1,$ be coordinates in two-dimensional Minkowski
space. Let us examine the Nambu action for a bosonic string
$$
S=-\frac {1} {l ^ 2} \, \int \,\sqrt{-g} \, d^2 \theta=\int \, d\tau \,
 {\cal L} \eqno(2.1)
$$
Here $\theta^a=(\tau ,\,\phi )$ are the parameters of the world sheet of the
string and
$$
g = Det \, g _ {ab} \, ,  \qquad g _ {ab} = \eta _ {\mu\nu} \,
\frac {\partial X^{\mu}} {\partial\theta^a} \,
\frac {\partial X ^ {\nu}} {\partial\theta^b}
$$
The parameter $\tau $ is timelike and $\phi $ is spatial. The partial
derivatives $\partial /\partial \tau $ and $\partial /\partial \phi $ will
be denoted below by an overdot and a prime, respectively. It is easy to show
that the generalized momenta $\pi _\mu =\partial {\cal L}/\partial X^\mu $
satisfy the conditions
$$
{\cal E} = \frac {l ^ 2} {2} \, \pi _ {\mu} \, \pi ^ {\mu} + \frac {1} {2 \, l ^ 2} \,
{X ^ {\mu}} ^ {\prime} \,
X ' _ {\mu} \approx 0 \,,
$$
$$
{\cal P} = {X ^ {\mu}} ^ {\prime} \, \pi _ {\mu} \approx 0 \eqno(2.2)
$$
The quantities ${\cal E}(\phi )$ and ${\cal P}(\phi )$ exhaust all
the first class constraints. The Hamiltonian of the system
$$
{ \cal H} = \int \, d\phi \,\pi _ {\mu} \, {\dot {\phi}} ^ {\mu} - {\cal L} = 0
$$
is also equal to zero.
For this reason, following Dirac, we must employ a generalized Hamiltonian
which is an arbitrary linear combination of the first class constraints (2.2)
$$
{\cal H} _ T = \int \, d\phi \, (v \, {\cal P} + w \, {\cal E}) \eqno(2.3)
$$
The equations of motion can be obtained from the variational principle
$$
\delta \, S = \delta \, \{\,\int \, d\tau \, (\,\int \, d\phi \,\pi _ {\mu} \,
{ \dot {X}} ^ {\mu} - {\cal H} _ T) \, \}=0           \eqno(2.4)
$$
In the case of an open string, when the parameter $\phi $ varies from 0 to
$\pi $, the variational principle (2.4) gives, besides the equations of
motion, the boundary conditions
$$
( v \,\pi _ {\mu} + w \,\frac {1} {l ^ 2} \, X ' _ {\mu}\,) \, \vert _ {\phi =
 0, \, \pi} = 0      \eqno(2.5)
$$
which ordinarily are replaced by the conditions
$$
v\vert _ {\phi = 0, \, \pi} = 0 \,, \qquad
X ' _ {\mu} \vert _ {\phi = 0, \, \pi} = 0     \eqno(2.6)
$$

For a closed string instead of the boundary condition there
is the periodicity condition.

Let us study an open string next.

The first step in the quantization process is to postulate the commutation
relations for the generalized coordinates and momenta:
$$
[ \, X ^ {\mu} (\phi), \, \pi ^ {\nu} (\phi ') \,] = i \,\eta ^ {\mu\nu} \, \delta (\phi-\phi ')
   \eqno(2.7)
$$
The commutation relations (2.7) and the boundary conditions (2.6) are
satisfied if
$$
X ^ {\mu} (\phi) = \frac {l} {\sqrt {\pi}} \, \left (x ^ {\mu} + i \,
\sum _ {n\neq 0} \, \frac {1} {n} \, \alpha ^ {\mu} _ n \,\cos {n\phi} \right) \, ,
$$
$$
\pi ^ {\mu} (\phi) = \frac {1} {\sqrt {\pi} \, l} \, \sum _ {n} \,
\alpha ^ {\mu} _ n \,\cos {n\phi} \,       \eqno(2.8)
$$
and the elements $(x^\mu ,$ $\alpha _n^\mu )\,$ satisfy the commutation
relations
$$
[ \, x ^ {\mu}, \, \alpha ^ {\nu} _ n \,] = i \,\eta ^ {\mu\nu} \, \delta _ n \,, \qquad
[ \, x ^ {\mu}, \, x ^ {\nu}\,] = 0 \,,
$$
$$
[ \, \alpha ^ {\mu} _ m, \, \alpha ^ {\nu} _ n \,] =
 m \,\eta ^ {\mu\nu} \, \delta _ {m + n}       \eqno(2.9)
$$
Since the quantities (2.8) are real,
$$
x ^ {\mu +} = x ^ {\mu} \, , \qquad      \alpha ^ {\mu +} _ n =
\alpha ^ {\mu} _ {-n}      \eqno(2.10)
$$
The constraints (2.2) can be represented as
$$
( {\cal E} \pm {\cal P}) (\phi) = \frac {1} {2} \, (\xi ^ {\mu} _ {\pm} (\phi)) ^ 2 \,,
   \eqno(2.11)
$$
where
$$
\xi ^ {\mu} _ {\pm} (\phi) = \frac {1} {\sqrt {\pi}} \, \sum _ n \,\alpha ^ {\mu} _ n \,
\exp\mp i \, n \,\phi      \eqno(2.12)
$$
Hence it is seen that ${\cal E}-{\cal P}$ differs from ${\cal E}+{\cal P}$
by the replacement $\phi $ by $-\phi $. This simplifies the analysis, since
on the interval $-\pi \le \phi \le \pi $ the quantity ${\cal E}+{\cal P}$
contains all information about the quantities ${\cal E}\pm {\cal P}$ on the
interval $0\le \phi \le \pi $. Therefore the Fourier components
 $$
 L_n = \frac {1} {2} \, \int ^ {\pi} _ {-\pi} \, d\phi \, ({\cal E} +
 {\cal P}) \, \exp \, i \, n \,\phi     \eqno(2.13)
 $$
are equivalent to the set of quantities (2.2) for $0\le \phi \le \pi $.
According to Eqs. (2.11)-(2.13), we have
$$
L_n = \frac {1} {2} \,: \,\sum _ m \,\alpha ^ {\mu} _ {n-m} \, \alpha _ {\mu \, m} \,:
      \eqno(2.14)
$$
The sense of the ordering operation in Eq. (2.14) is determined by the
quantization method.

Let us also write out expressions for the momentum and angular momentum of a
string:
$$
P ^ {\mu} = \int ^ {\pi} _ 0 \, d\phi \,\pi ^ {\mu} \, , \qquad
 J ^ {\mu\nu} = \int ^ {\pi} _ 0 \, d\phi \, (X ^ {\mu} \, \pi ^ {\nu} -
 X ^ {\nu} \, \pi ^ {\mu})       \eqno(2.15)
 $$
With the help of Eqs. (2.6) and (2.7) we immediately verify that
$$
[ \, P ^ {\mu}, \, {\cal H} _ T \,] = 0 \,, \qquad [\, J ^ {\mu\nu}, \, {\cal H} _ T \,] = 0
$$
This means that the momentum and angular momentum of the string are
conserved.

In the currently conventional quantization the ground state $\left|
0\right\rangle \,$ satisfies the conditions
 $$
 \alpha ^ {\mu} _ m \,\vert \, 0 \,\rangle = 0 \,, \qquad m\geq 0 \eqno (2.16)
$$
The whole space of states has the orthogonal basis
$$
\alpha ^ {\mu _ 1} _ {m _ 1} \, \ldots \,\alpha ^ {\mu _ s} _ {m _ s} \, \vert \, 0 \,\rangle \,,
 \qquad m _ i < 0        \eqno(2.17)
$$
Therefore all $\alpha_m^{\mu}$ are linear operators in the whole space of
states. From Eqs. (2.9) and (2.6) it follows that the metric in the state
space (2.17) is indefinite. The ordering in Eq. (2.14) means that the
operators $\alpha _m^\mu $ with $m<0$ are arranged to the left of all
operators $\alpha _n^\mu $ for $n\ge 0$. With this ordering the commutators
of Virasoro operators contain anomalies
$$
[ \, L _ n, \, L _ m \,] = (n-m) \, L _ {n + m} + \frac {1} {12} \, D (n ^ 3-n)
       \eqno(2.18)
$$
Here $D$ is the dimension of the $x$-space, which in our case is 2.
Therefore the maximum that can be achieved is annihilation of the operators
 $L_n$ with $n\ge 0$. As a result the theory is consistent only for $D=26$.
 A detailed study of the problems arising under the quantization (2.16) can be
found in Ref. [4].

We shall now present the path proposed here for quantization of a
two-dimensional string that leads to a self-consistent theory of a string in
a space of two dimensions. Our method of quantization of a string is similar
to Dirac's method of quantization of the electromagnetic field (see Ref. [5]
, and also Appendix).

Let
$$
x_{\pm} = x ^ 0\pm x ^ 1 \,, \qquad \alpha ^ {(\pm)} _ m =
\alpha ^ 0 _ m\pm\alpha ^ 1 _ m
  \eqno (2.19)
$$
From Eq. (2.9) we obtain:
$$
[ \alpha ^ {(+)} _ m, \, \alpha ^ {(+)} _ n \,] =
[ \alpha ^ {(-)} _ m, \, \alpha ^ {(-)} _ n \,] = 0 \,, \qquad
[ \alpha ^ {(+)} _ m, \, \alpha ^ {(-)} _ n \,] =-2m \,\delta _ {m + n}
$$
$$
[ x _ +, \, x _ - \,] = 0 \,, \qquad \ [x _ +, \, \alpha ^ {(+)} _ n \,] =
[ x _ -, \, \alpha ^ {(-)} _ n \,] = 0 \,,
$$
$$
[ x _ +, \, \alpha ^ {(-)} _ n \,] =-2i \,\delta _ n \,,      \qquad \
 [x _ -, \, \alpha ^ {(+)} _ n \,] =-2i \,\delta _ n        \eqno (2.20)
 $$
Let us write the Virasoro operators in the variables $\alpha ^{(\pm )}$:
$$
L _ n = -\frac {1} {2} \,: \,\sum _ m \,\alpha ^{(+)}_{n-m} \, \alpha ^{(-)} _ m \,:
  \eqno (2.21)
$$

By definition, the ordering operation in Eq. (2.21) means that either
the elements $\alpha ^{(+)}$ are arranged to the left of all elements
 $\alpha ^{(-)}$ or the elements $\alpha ^{(-)}$ are arranged to the left of
all elements $\alpha ^{(+)}$. Both orderings are equivalent. Indeed,
$$
\sum _ m \,\alpha ^ {(-)} _ m \,\alpha ^ {(+)} _ m =
\sum _ m \,\alpha ^ {(+)} _ m \,\alpha ^ {(-)} _ m + 2 \,\sum _ m \, m \,,
$$
It can be assumed that the last sum is zero, since it can be written as
$\zeta (-1)-\zeta (-1)$, where $\zeta (s)$ is the Riemann zeta-function. It
is known that the zeta-function
 $$
 \zeta (s) \equiv\sum ^ {\infty} _ {n = 1} \, n ^ {-s}
 $$
possesses a unique analytical continuation to the point $s=-1$ and $\zeta
(-1)=-1/12$.

For definiteness, let us choose the same ordering as in Eq. (2.21).

According to Eq. (2.20), we have
$$
[ \, L _ m, \, \alpha ^ {(-)} _ n \,] =-n \,\alpha ^ {(-)} _ {m + n}     \eqno(2.22)
$$
One can see from Eqs. (2.20) and (2.22) that the weak equalities $\alpha
_n^{(-)}\approx 0$ and $L_n\approx 0$ are algebraically consistent. For this
reason, we determine the physical states as the states satisfying the
conditions
$$
\alpha ^ {(-)} _ n \,\vert \ \rangle = 0 \,, \qquad n = 0, \, \pm1, \, \ldots
        \eqno (2.23)
$$
It follows immediately from Eqs. (2.23) and (2.21) that
$$
L _ n \,\vert \ \rangle = 0 \,, \qquad n = 0, \, \pm1, \, \ldots    \eqno(2.24)
$$
for any physical states. The equalities (2.24) mean that for the
quantization (2.23) the Virasoro algebra has no anomalies:
$$
[ L _ n, \, L _ m \,] = (n-m) \, L _ {n + m} \eqno (2.25)
$$
The latter formula can also be easily obtained by direct calculation of the
commutators, provided that the ordering is assumed to be the same as in Eq.
(2.21). The quantization conditions (2.23) are precisely analogous to the
quantization conditions (A.8) used by Dirac to quantize the electromagnetic
field [5].

We call attention to the fact that states of the form
$$
\alpha ^ {(+)} _ n \,\vert \ \rangle \,, \qquad n\neq 0     \eqno(2.25)
$$
are not considered in this theory, since these states do not satisfy the
conditions (2.23). For this reason, the matrix elements of the quantities
$\alpha _n^{(+)}$ with $n\ne 0$ with respect to the physical states (2.26)
cannot be calculated. Therefore the quantities $\alpha _n^{(+)}$ with $n\ne
0 $ cannot be operators in the space of physical states. Hence it follows
that observables cannot depend on the elements $\alpha _n^{(+)}$ with $n\ne
0 $. In other words, observables must commute with all operators $\alpha
_n^{(-)}$. According to Eq. (2.20), there are two such quantities: $x_{-}$
and $p_{+}$ ($p_{\pm }\equiv \alpha _0^{(\pm )}$). Both are real.

Thus we can see that the quantities $\alpha _n^\mu $ with $n\ne 0$ are not,
generally speaking, linear operators in state space in the conventional
sense. Here we adhere to the concept formulated and applied by Dirac in Ref.
[5]. According to this concept, in quantum field theory linear operators
acting in certain linear spaces are replaced by so-called $q$ numbers,
which form an associative noncommutative algebra with an involution over the
complex numbers. Here we shall formulate Dirac's concept using the
conventional mathematical terminology.

Let ${\cal A}$ be an associative noncommutative involutive algebra with an
identity over the complex numbers. Associativity means that for any elements
$u,\,v,$ and $w$ of the algebra ${\cal A}$ and any number $c$ the
equalities
 $$
 (u \, v) \, w = u \, (v \, w) \, , \qquad (cu) \, v = u \, (cv) = c (u \, v)
 $$
hold. The involution property of the algebra means that there exists a
mapping $u\mapsto u^{+}\,$ from ${\cal A}$ into ${\cal A}$ such that
$$
 (u ^ +) ^ + = u \,, \ \ \ (c _ 1u + c _ 2v) ^ + = {\bar {c}} _ 1u ^ ++ {\bar {c}} _ 2v ^ + \,,
  \ \ \ (u \, v) ^ + = v ^ + \, u ^ +
  $$
for any $u,\,v$ $\in {\cal A}$ and any numbers $c_1$ and $c_2$. An overbar
signifies complex conjugation. If $u^{+}=u$, the element $u$ is said to be
hermitian.

It is also assumed that the algebra ${\cal A}$ has a system of generators
 $\{\alpha _p\}$ for which all relations are limited by the form of the
commutators
$$
[ \alpha _ p, \, \alpha _ {p '}\,] = c _ {p \, p '}
$$
Here $c_{pp^{\prime }}$ are complex $c$-numbers (in the Dirac sense).

The definition of involutive algebras (or algebras with involution) and
other mathematical definitions presented here can be found in Refs. [6]
and [7].

Let $V$ be a vector space with elements $\left| \Lambda \right\rangle ,$
 $\left| \Sigma \right\rangle ,$ ... over the complex numbers and
 let $V^{+}$
 be the conjugated space, whose elements are denoted by $\left\langle \,
{}\right| $. There is a one-to-one correspondence between the elements of
the spaces $V$ and $V^{+}$ such that $c\left| \Lambda \right\rangle
\leftrightarrow \left\langle \Lambda \right| \bar{c}$.

For any two vectors $\left| \Lambda \right\rangle $ and $\left| \Sigma
\right\rangle $ there exist two complex mutually-conjugated $c$-number
 quantities
$\left\langle \Lambda \mid \Sigma \right\rangle $ and $\left\langle \Sigma
\mid \Lambda \right\rangle $. It is assumed that in the space $V$ there
exists a basis $\{\left| \Lambda \right\rangle \}$ such that
$$
\langle\,\Lambda\,\vert\,\Sigma\,\rangle = \delta_{\Lambda\Sigma} \eqno(2.27)
$$
If the indices $\Lambda $ and $\Sigma $ run a continuous set, then
in Eq. (2.27) $\delta _{\Lambda \Sigma }$ must be interpreted as a
delta-function. The space $V$ is the space of physical states of the theory.

Let ${\cal B}\subset {\cal A}$ be a noncommutative involutive subalgebra
with the identity element. The elements of the subalgebra ${\cal B}$ are
linear operators in the spaces $V$ and $V^{+}$ and, as usual,
$$
(\,u\,\vert\,\Lambda\,\rangle\,)^+ =\langle\,\Lambda\,\vert\, u^+\,,
 \qquad u\in {\cal B}
$$
The observables correspond to certain hermitian elements from ${\cal B}$. If
$u\in {\cal A}$ and $u\notin {\cal B}$, then the action of the element $u$
on vectors from the spaces $V$ and $V^{+}$, generally speaking, is not
defined. This distinguishes the Dirac theory from the conventional quantum
field theory.

In the theories under study all vectors of the space $V$ are, ordinarily,
annihilated by a series of operators of the subalgebra ${\cal B}$. Therefore
the conditions
$$
u_N\,\vert \ \rangle=0\,, \qquad u_{N '}\,\vert \ \rangle = 0 \,,\, \ldots\,,
\ \qquad \vert \ \rangle\in V     \eqno(2.28)
$$
take place. The indices $N,$ $N^{\prime },$ ... in Eq. (2.28) run a
certain set $J$ of indices. The conditions (2.28) must be algebraically
 consistent, i.e. the equalities
$$
[\,u_N,\, u_{N '}\,]=\sum_{N} \,\kappa_{NN ',\, N^{\prime \prime}}
\, u _ {N^{\prime \prime}}
$$
where $N,$ $N^{\prime },$ $N^{\prime \prime }\in J$ and $k_{NN^{\prime
},N^{\prime \prime }}$ can be any elements of the algebra ${\cal A}$, must
hold. Evidently, the operators $u_N$ in Eq. (2.28) do not include the
identity operator. We denote by ${\cal N\subset B}$ the subalgebra
without identity with the generators $\{u_N\}$, where $N\in J$.
Thus subalgebra ${\cal N}$ annihilates the space of physical states $V$.

Let us now examine the set of elements of the form $ru$, where $r\in {\cal A}
$ and $u\in {\cal N}$. We denote this set as ${\cal N}^{\prime }$. It is
evident from the definition that ${\cal N}^{\prime }$ is a left
${\cal A}$-module and a subalgebra in ${\cal A}$,
but ${\cal N}^{\prime }$ is not a
subalgebra in ${\cal B}$. Nonetheless,
 the action of the subalgebra ${\cal N}^{\prime }$ on the
 space $V$ is defined since it is trivial: ${\cal N}^{\prime }$
 annihilates the space $V$. We note that the
 commutant $[{\cal N}^{\prime },$ ${\cal N}^{\prime }]$ is contained
 in ${\cal N}^{\prime }$.
Indeed, if $r,\,s\in {\cal A}$ and $u,\,v\in {\cal N}$, then
$$
[\, ru,\, sv\,]=\{[\, ru, \, s \,] \, v + s \, [\, r, \, v \,] \, u +
sr \, [\, u, \, v \,] \}\in {\cal N} '
$$
since $[u,\,v]\in {\cal N}$ . Then the conditions
 ${\cal N}^{\prime }V=0$ are algebraically consistent.

Concrete theories can also contain other elements of the algebra ${\cal A}$,
which are not contained in either ${\cal B}$ or ${\cal N}^{\prime }$ and are
linear operators on the space $V$.

A distinguishing feature of the Dirac theory is the fact that nonphysical
state vectors that do not satisfy the conditions (2.28) are not considered
 in it. Moreover, in the Dirac theory an indefinite metrics
in the state space is absent. This circumstance can radically
alter the theory.

Let us return to the discussion of string theory. In the theory proposed
here for a two-dimensional string the algebra ${\cal A}$ has generators
$\{x_{\pm },\alpha _m^{(\pm )}\}$, while the subalgebras ${\cal B}$ and
${\cal N\,}$ have generators $\{x_{-},p_{+},\alpha _m^{(-)}\}$ and $\{\alpha
_m^{(-)}\},$ respectively. The Virasoro operators $L_n$ are contained in the
subalgebra ${\cal N}^{\prime }$. We note that the algebra of operators $L_n$
is an involutive subalgebra in ${\cal N}^{\prime }$, and since
 $L_n^{+}=L_{-n},$ the action of the operators $L_n$ is defined in
 both spaces $V$ and $V^{+}$.

From the definitions (2.15) we obtain the following formulas:
$$
( \exp \, i\omega \, J ^{01}) \, \alpha ^ {(\pm)} _ m \, (\exp-i\omega \, J ^ {01}) =
(\exp\pm\omega) \, \alpha ^ {(\pm)} _ m \,,
$$
$$
( \exp \, i\omega \, J ^ {01}) \, x_{\pm}\, (\exp-i\omega \, J ^ {01}) =
(\exp\pm\omega) \, x _ {\pm} \eqno (2.29)
$$
and
$$
( \exp \, ia _ {\mu} P ^ {\mu}) \, x_{\pm}\, (\exp-ia _ {\mu} P ^ {\mu}) =
X _ {\pm} + \frac {\sqrt {\pi}} {l} \, a _ {\pm} \,
$$
$$
( \exp \, ia _ {\mu} P ^ {\mu}) \, \alpha^{(\pm)}_m \, (\exp-ia _ {\mu} P ^ {\mu}) =
\alpha ^ {(\pm)} _ m \eqno (2.30)
$$
Here $\omega $ and $a^\mu $ are arbitrary real numbers. It is evident from
Eqs. (2.29) and (2.30) that translations and Lorentz transformations
conserve the condition (2.23).

Both observables $x_{-}$ and $p_{+}=\alpha _0^{+}$ are real, and $[x_{-},$
$p_{+}]=-2i$. For this reason, we assume that the physical states are
eigenstates of the operator $p_{+}$:
 $$
 p_+ \, \vert \, k \,\rangle = 2k \,\vert \, k \,\rangle \eqno (2.31)
 $$
Here $k$ is a continuous real parameter. According to Eq. (2.29)
$$
p_+ \, (\exp-i\omega \, J ^ {01}) = (\exp \,\omega)\,
 (\exp-i\omega \, J ^ {01}) \, p _ +
  \eqno (2.32)
  $$
Let us formally act with the equality (2.32) on the state $\left|
k\right\rangle $. As a result of Eq. (2.31) we obtain
$$
p_+ \, (\exp-i\omega \, J ^ {01}) \, \vert \, k \,\rangle = 2k \, e ^ {\omega} \,
( \exp-i\omega \, J ^ {01}) \, \vert \, k \,\rangle \eqno (2.33)
$$
The last equality makes it possible to determine the action of the operators
$(\exp -i\omega J^{01})$ on the physical states as follows:
$$
( \exp-i\omega \, J ^ {01}) \, \vert \, k \,\rangle =
f_ {\omega} \, \vert \, e ^ {\omega} k \,\rangle
 \eqno (2.34)
 $$
Here $f_\omega $ is a complex number different from zero. If the scalar
product on physical state vectors is defined in a Lorentz-invariant manner
as
$$
\langle \, k \,\vert \, k ' \,\rangle = k \,\delta (k-k ')
$$
then $\left| f_\omega \right| =1$. From Eq. (2.34) it is evident that one
can assume
$$
k> 0 \eqno (2.35)
$$
The angular momentum operator can be represented in the form
$$
J ^ {01} = \frac {1} {2} \, (x _ -p _ + -x _ + p _ -) +
\frac {i} {4} \, \sum _ {n\neq 0} \, \frac {1} {n} \,
(\alpha ^ {(-)} _ n \,
\alpha ^ {(+)} _ {-n} -\alpha ^ {(+)} _ n \,\alpha ^ {(-)} _ {-n})\,
 \eqno (2.36)
$$
We can see that although the expression (2.36) does not belong to either the
subalgebra ${\cal B}$ or the subalgebra ${\cal N}^{\prime }$, the action of
the quantities $(\exp i\omega J^{01})$ on the space of physical states is
nonetheless correctly determined.

According to Eqs. (2.8) and (2.15)
$$
P ^ {\mu} = \frac {\sqrt {\pi}} {l} \, \alpha ^ {\mu} _ 0 =
\frac {\sqrt {\pi}} {2 \, l} \,
\{ (\delta^{\mu}_0 + \delta^{\mu} _ 1) \, p _ ++
 (\delta^{\mu}_0-\delta^{\mu} _ 1) \, p _ - \}
$$
Therefore from (2.23) and (2.31) we obtain:
$$
P^{\mu} \, \vert\, k\,\rangle =
\frac{\sqrt {\pi}} {l}\, k ^ {\mu} \, \vert \, k \,\rangle \,,
 \qquad k^{\mu} = (k, \, k) \eqno (2.37)
 $$

Thus, as a result of the above-described procedure of quantization of
two-dimensional string there arises a system similar to a massless chiral
quantum particle in two-dimensional space-time.

\centerline {}
\centerline {\bf {3. The conclusion}}
\centerline {}

Let us note the differences of the main properties of string theory
quantized in the conventional manner from those of the string theory
proposed in the present paper. In the conventional quantization there exists
a state which is invariant under Lorentz transformations. This state is the
ground state. In this respect the conventional string theory is similar to
the standard quantum field theory of point objects. In such field theories
the ground state usually is Lorentz-invariant. Conversely, in our
approach there does not exist a state that is invariant under Lorentz
transformations. For this reason, the quantum-string theory proposed above
is analogous to a quantum theory of a single relativistic particle. Once
again there does not exist a Lorentz-invariant quantum state of a single
relativistic particle. In order for a Lorentz-invariant state to exist in
our theory we would have to introduce a string field and second-quantize the
string field. In such a theory the ground state would be Lorentz-invariant,
since there would be no real strings in the ground state.

In closing, we note that the quantization method proposed here can be
applied to a $D$-dimensional string. This assertion is based on the fact
that in string theory there exists an infinite set of so-called {\it DDF}-
operators Ref. [4] which commute with all Virasoro operators. The
{\it DDF}-operators describe almost all (with the exception of the total momentum of
the string) physical degrees of freedom of the string. The independence of
Virasoro operators from {\it DDF }-operators means that Virasoro
operators can be put into the form (2.21), after which the quantization
scheme which we have proposed above can be applied. However, the theory is
much more complicated in the $D$-dimensional case because there exists an
infinite set of physical degrees of freedom, contained in the
{\it DDF}-operators.

\centerline {}
\centerline {\sc { Appendix}}
\centerline {}

We shall describe the quantization of a free electromagnetic field
 following Dirac's ideology Ref. [5],
which is formulated in Sec. 2. The
quantization which we propose for a two-dimensional string is performed in
accordance with Dirac's ideology.

The quantization of an electromagnetic field is presented in the form
$$
A_{\mu}(x) = \int \,\frac {d ^ 3k} {(2\pi) ^ 3} \, \frac {1} {\sqrt {2k ^ 0}} \,
\{a _ {\mu} ({\vec {k}}) \, e ^ {ikx} + a ^ + _ {\mu}
 ({\vec {k}}) \, e ^ {-ikx} \}
  \eqno (A1)
$$
Here $\mu ,\,\nu ,$ .... = 0, 1, 2, 3, \ $kx\equiv k_\mu x^\mu =-k^0x^0+
{\vec k}\cdot {\vec x},$ \ $k^0=\left|{\vec k}\right| $ and
 $\{a_\mu ({\vec k}),$ $a_\mu^{+}({\vec k})\}\,\,$
 are some generators of an associative involutive algebra
 ${\cal A}$ with an identity element (see Sec. 2). The nonzero commutation
relations between these generators have the form
$$
[ \, a _ {\mu} ({\vec {k}}), \, a ^ + _ {\mu} ({\vec {p}})\,] =
(2\pi) ^ 3 \,\eta _ {\mu\nu} \, \delta ^ {(3)} ({\vec {k}} - {\vec {p}})
  \eqno (A2)
 $$
One can see from the expansion (A1) that the set of elements $\partial _\mu
A^\mu (x)$ is linearly equivalent to the set of
elements $k^\mu a_\mu ({\vec k})$
and $k^\mu a_\mu ^{+}({\vec k})\,$ from the algebra ${\cal A}$. Let
 $a_i^T({\vec k})$ be two independent elements
  (for fixed ${\vec k}$) satisfying
the conditions
$$
\sum ^ 3 _ {i = 1} \, k _ i \, a ^ T _ i ({\vec {k}}) = 0 \,,
$$
$$
[ \, a ^ T _ i ({\vec {k}}), \, a ^ {T +} _ j ({\vec {p}})\,] =
(2\pi) ^ 3 \,\left (\delta _ {ij} -\frac {k _ ik _ j} {{\vec {k}} ^ 2} \right) \,
\delta ^ {(3)} ({\vec {k}} - {\vec {p}}) \eqno (A3)
$$
Eqs.(A1) and (A2) imply the following commutation relations
 \ $ (F _ {\mu\nu} = \partial _ {\mu} A _ {\nu} -\partial _ {\nu} A _ {\mu}) $:
 $$
[ \, F _ {\mu\nu}(x),\,k ^ {\lambda} \, a _ {\lambda} ({\vec {k}})\,] =
[ \, F _ {\mu\nu}(x),\, k ^ {\lambda} \, a ^ +_ {\lambda} ({\vec {k}})\,]=0 \,,
  \eqno (A4)
  $$
  $$
  [\, k ^ {\mu} \, a _ {\mu} ({\vec {k}}), \, p ^ {\nu} \, a ^ + _ {\nu}
  ({\vec {p}})\,] = 0
    \eqno (A5)
    $$
It is obvious, that
$$
[ \, a ^ T _ i, \, k ^ {\mu} \, a _ {\mu} ({\vec {k}})\,] =
[ \, a ^ T _ i, \, k ^ {\mu} \, a ^ + _ {\mu} ({\vec {k}})\,] = 0 \eqno (A6)
$$
Dirac quantization presupposes that the conditions
$$
A ^ T _ i ({\vec {k}}) \, \vert \, 0 \,\rangle = 0    \eqno (A7)
$$
are imposed on the ground state and the conditions
$$
k^ {\mu} \, a _ {\mu} ({\vec {k}}) \, \vert \ \rangle = 0 \,, \qquad
k^ {\mu} \, a ^ + _ {\mu} ({\vec {k}}) \, \vert \ \rangle = 0 \eqno (A8)
$$
are imposed on all states. As a result of Eqs. (A5) and (A6) the
conditions (A7) and (A8) are algebraically consistent. The states
satisfying the conditions (A8) are called physical. The Fock space of all
physical states is constructed with the help of the creation operators
 $a_i^{T+}({\vec k})$ from the ground state satisfying the conditions
  (A7) and (A8). As a result of Eq. (A6) any state of the Fock space
  constructed satisfies the conditions (A8). Following the terminology
   introduced in Sec.
 2, this Fock space is designated by the symbol $V$, the set of elements
 $\{a_i^T,$ $a_i^{T+},$ $k^\mu a_\mu ({\vec k}),$ $k^\mu a_\mu ^{+}({\vec k})\}$
is a system of generators of the subalgebra ${\cal B}$ and the set of
elements $\{k_\mu a_\mu ({\vec k}),\,k^\mu a_\mu ^{+}({\vec k})\}$ is a system
of generators of the subalgebra ${\cal N}$.

Let $k_{-}^\mu =(-k^0,{\vec k})$. We find from Eq. (A2)
$$
[ \, k_{-}^{\mu} a _ {\mu} ({\vec {k}}), \, p ^ {\nu} \, a ^ + _ {\nu} ({\vec {p}})\,] =
2 \, {\vec {k}} ^ 2 \, (2\pi) ^ 2 \,\delta ^ {(3)} ({\vec {k}} - {\vec {p}})
  \eqno (A9)
$$
The relations (A4) and (A9) mean that the observables $F_{\mu \nu }$ do
not depend on the generators $\{k_{-}^\mu a_\mu ({\vec k}),$
 $k_{-}^\mu a_\mu ^{+}({\vec k})\}$, but
 rather they are linear combinations of the generators of
the subalgebra ${\cal B}$. Therefore all matrix elements of the form
$\left\langle \Lambda \right| F_{\mu \nu }\left| \Sigma \right\rangle $,
where $\left| \Lambda \right\rangle ,$ $\left| \Sigma \right\rangle \in V$,
are determined.

We note that as a result of Eqs. (A3) and (A7) the scalar product in the
space $V$ is positive-defined provided that $\left\langle 0\mid
0\right\rangle =1$. We call attention to the fact that the action of the
generators $k_{-}^\mu a_\mu ({\vec k})$ and $k_{-}^\mu a_\mu ^{+}({\vec k})$
on the physical states is not determined in Dirac quantization, and
therefore these generators of the algebra ${\cal A}$ are not linear
operators in the space of physical states $V$.

In closing, we call attention to an analogy between the generators
$\{k_{-}^\mu a_\mu ({\vec k}),\,k_{-}^\mu a_\mu ^{+}({\vec k})\}$ and $\{k^\mu
a_\mu ({\vec k}),$ $k^\mu a_\mu ^{+}({\vec k})\}$ in quantum electrodynamics
and the generators $\{\alpha _n^{(+)}\}$ and $\{\alpha _n^{(-)}\}\,$ in
string theory, respectively.

\centerline {}
\centerline {REFERENCES}
\centerline {}

\begin {itemize}
\item [1].
R.Jackiw. Solutions to a Quantal Gravity-Matter Field Theory
On a Line. Preprint gr-qc/9612052, 18 Dec 1996.
\end {itemize}
\vspace {-4mm}
 \begin {itemize}
\item [2]. D. Gangemi and R. Jackiw, Phys. Lett. {\bf {B337}},
271 (1994); Phys. Rev. {\bf {D50}}, 3913 (1994); D. Amati, S. Elitzur and E.
Rabinovici, Nucl. Phys. {\bf {B418}}, 45 (1994); D. Louis-Martinez, J
Gegenberg and G. Kunstatter, Phys. Lett. {\bf {B321}}, 193 (1994); E.
Benedict, Phys. Lett. {\bf {B340}}, 43 (1994); T. Strobl, Phys. Rev. {\bf
{D50}}, 7346 (1994). \end {itemize}
\vspace {-4mm}
\begin {itemize}
\item [3].
C.Callan, S.Giddings, J.Harvey, A.Strominger, Phys. Rev. {\bf {D45}},
 1005 (1992)
\end {itemize}
\vspace {-4mm}
\begin {itemize}
\item [4].
M.B. Green, J.H. Schwarz, E. Witten. Superstring Theory. Cambridge
University Press, 1987
\end {itemize}
\vspace {-4mm}
\begin {itemize}
\item [5]. P.A.M. Dirac. Lectures on quantum field theory. Yeshiva University,
 New York, 1967
\end {itemize}
\vspace {-4mm}
\begin {itemize}
\item [6]. Serge Lang. Algebra.
Columbia University, New York. 1965
\end {itemize}
\vspace {-4mm}
\begin {itemize}
\item [7]. N. Bourbaki. Theories spectrales. Hermann. 1967
\end {itemize}

\end{document}